\documentclass[fleqn,twoside]{article}
\usepackage{espcrc2}
\usepackage{graphicx}
\usepackage[figuresright]{rotating}
\newcommand{\ba}{$\tt BABAYAGA$}
\newcommand{\bh}{$\tt BHWIDE$}
\newcommand{\labs}{$\tt LABSPV$}
\newcommand{\oal}{${\cal O}(\alpha)$}
\title{Matrix elements and Parton Shower in the event generator $\tt BABAYAGA$}
\author{G. Balossini\address[uni]{Dipartimento di Fisica Nucleare e Teorica, 
Universit\`a di Pavia, via A. Bassi 6, Pavia (Italy)},
C.M. Carloni Calame\address[infn]{Istituto Nazionale di Fisica Nucleare,
Sezione di Pavia, via A. Bassi 6, Pavia (Italy)}\addressmark[uni],
G. Montagna\addressmark[uni]\addressmark[infn],
O. Nicrosini\addressmark[infn],
F. Piccinini\addressmark[infn]\addressmark[uni]}
\begin{document}
\begin{abstract}
\noindent
A new version of the event generator \ba\ is presented, which is based
on an original matching of the Parton Shower approach with the complete
exact \oal\ matrix element for the inclusion of the QED radiative
corrections to the Bhabha process at flavour factories. The
theoretical accuracy of the improved generator is conservatively
estimated to be 0.2\%, by comparison with independent calculations. The
generator is a useful tool for precise luminosity determination at
flavour factories, for center of mass energies below 10 GeV.
\vspace{1pc}
\end{abstract}
\maketitle
\section{Introduction}
\label{intro}
The precise determination of the machine luminosity is an important ingredient
for the successful achievement of the physics programme at the $e^+e^-$
colliders running with center of mass energy in the range of the low lying
hadronic resonances.

One of the most important challenges is the precise measure of
the $R$ ratio, by means of the energy scan or the radiative return method.
The aim is to reduce the theoretical error on the hadronic contribution to the
vacuum polarization, which reflects on the error of the anomalous magnetic
moment of the muon $a_\mu$ and the QED coupling constant at the $Z$ peak
$\alpha_{QED}(M_Z^2)$~\cite{fj}. The precise measure of $R$ will
give a stringent test of the Standard Model predictions.

In order to achieve a precise determination of the collider luminosity,
high-precision calculations of the QED  processes $e^+ e^- \to e^+
e^-, \mu^+ \mu^-, \gamma\gamma$, and relative Monte 
Carlo generators, are required. The large-angle Bhabha process is, in
particular, of major interest for its large cross section and
its clean experimental signature.

The event generator \ba\ was developed~\cite{ournpb,myplb} to this end and
it is widely
used by experimental collaborations.
The event generators $\tt MCGPJ$~\cite{mcgpj}, $\tt
BHAGENF$~\cite{bhagenf} and $\tt BHWIDE$~\cite{bhwide}
are also employed by experimental collaborations.

The theoretical precision of the \ba\ event generator for the Bhabha
process was estimated at the 0.5\% level~\cite{ournpb}. The need of
an improvement in the calculation of radiative corrections within
\ba\ emerged in the last years, for a number of reasons. First, the
total luminosity error
quoted by KLOE is presently 0.6\%~\cite{kloe}, where the dominant
source of uncertainty comes from theory, i.e. from the 0.5\% physical
precision of the \ba\ generator. Secondly, the measurement
of the hadronic cross section 
in the $\pi^+ \pi^-$ channel at VEPP-2M has achieved 
a total systematic uncertainty of 0.6\%, requiring an assessment of the
collider luminosity at the level of 0.1\%. Last but not least, precision measurements
of $R$ trough radiative return at
KEK-B and PEP-II are already performed or foreseen in the near future.

Here, we describe the main features of a high-precision
calculation of photonic radiative corrections to the Bhabha process,
in order to improve the theoretical formulation of the original \ba\
generator up to $\cal{O}$(0.1\%). The
approach is based on the matching of exact next-to-leading (\oal)
order corrections
with resummation through all orders of $\alpha$ of the leading
contributions due to multiple radiation, taken into account
according to a QED Parton Shower (PS).

\section{Matching exact \oal\ matrix element and Parton Shower}
\label{sectionmatching}
The original version of \ba\ ($\tt 3.5$) is based on a PS in QED which allows
to calculate the photonic radiative corrections to the Bhabha
process. The PS is a numerical algorithm which exactly solves the DGLAP
equation for the electron Structure Function (SF) in QED, allowing to
include essentially the leading
logarithmic (LL) corrections to the cross section up to all orders of
$\alpha$. The main advantage
of the PS approach is that, thanks to its Monte Carlo nature, the
events can be generated exclusively, i.e. all the
momenta of the final state particles (fermions and an indefinite
number of photons) can be reconstructed. The PS can be improved
to account also for interference effects between initial and final
state radiation, allowing thus for a more accurate description of the
radiative events~\cite{myplb}.

The corrected differential cross section in the PS approach can be written (in
a simplified form for the sake of clarity) as
\begin{eqnarray}
&&d\sigma^\infty=\Pi(Q^2,\varepsilon)F_{SV}\Big\{d\sigma_0+\sum_{n=1}^\infty
\frac{d\hat\sigma_0}{n!}\times\nonumber\\
&&\prod_{i=1}^n\Big[\frac{\alpha}{2\pi}P(x_i)I(k_i)dx_idc_i
\theta(x_i-\varepsilon)\ F_{i,H}\Big]
\Big\}\label{sigma}
\end{eqnarray}
where $F_{SV}$ and $F_{i,H}$ are identically equal to 1 in the pure PS
case, $\Pi(Q^2,\varepsilon)=\exp(-\frac{\alpha}{2\pi}I_+L)$ is the
Sudakov form factor ($Q^2={st}/{u}$ is the scale in the collinear
logarithm $L=\log({Q^2}/{m^2_e})-1$ and
$I_+=\int_\varepsilon^1P(y)dy$), $P(x)$ is the Altarelli-Parisi
splitting function, $x_i$ is the energy fraction carried by the
$i^{th}$ photon, $\varepsilon$ is an infrared separator (the
integrated cross section is independent from its value)
and $I(k_i)$
is the photon angular spectrum function including initial-final state
interference. In Eq.~(\ref{sigma}) the photonic radiative corrections are
resummed up to all orders, in the LL accuracy.

The expansion at \oal\ of Eq.~(\ref{sigma}) is
\begin{eqnarray}
&&d\sigma^{\alpha,PS}=[\Pi(Q^2,\varepsilon)]_{{\cal O}(\alpha)}d\sigma_0+
\frac{\alpha}{2\pi}d\hat\sigma_0dxdc\times\nonumber\\
&&P(x)I(k)\theta(x-\varepsilon)=d\sigma_{SV}^{\alpha,PS}
+d\sigma_{H}^{\alpha,PS}\label{psoal}
\end{eqnarray}
where the subscript $SV$ defines the cross section with the emission
of a virtual or soft photon (with energy fraction lower than
$\varepsilon$) and $H$ defines the hard bremsstrahlung cross section (with
energy larger than $\varepsilon$). The PS error
is due to the fact that both the soft plus virtual and
hard part are approximated at the LL accuracy. By comparing the cross
section of Eq.~(\ref{psoal}) with the exact \oal\ cross section it
was possible to establish that the theoretical accuracy of \ba\ was at
$0.5\%$.

In order to go beyond the LL approximation and preserve the
resummation of the higher orders (h.o.) corrections, an appropriate procedure
must be devised, able to avoid the double counting of
the LL contributions and guarantee the independence of the cross
section from the $\varepsilon$ parameter. A solution to this
problem can be obtained by setting the factors $F_{SV}$ and $F_{i,H}$ to
\begin{eqnarray}
F_{SV}&=&1+\frac{d\sigma^{\alpha,ex}_{SV}-d\sigma^{\alpha,PS}_{SV}}{d\sigma_0}
\nonumber\\
F_{i,H}&=&1+\frac{d\sigma^{\alpha,ex}_{i,H}-d\sigma^{\alpha,PS}_{i,H}}
{d\sigma^{\alpha,PS}_{i,H}}\label{ffactors}
\end{eqnarray}
where $d\sigma_{SV}^{\alpha,ex}$ and $d\sigma_{H}^{\alpha,ex}$ are the
complete expressions for the soft plus virtual \oal\ cross
section~\cite{svlep1} and the real one photon emission cross section. By
construction, the factors of Eq.~(\ref{ffactors}) are infrared safe
quantities and they let the \oal\ expansion of $d\sigma^\infty$ in
Eq.~(\ref{sigma}) coincide with the exact \oal\ cross section. It is
worth noticing that correcting each single photon emission with a
factor $F_{i,H}$ is crucial to make the cross section independent from
the $\varepsilon$ parameter and that the correction is larger in those
phase space regions where the PS is more unreliable, typically where
the photon is hard and/or not collinear to one of the charged
particles. Notice that Eq.~(\ref{sigma}) is cast
in a completely differential form, so that events can be
generated exclusively as in the pure PS approach.

A more detailed discussion of the matching procedure, as well as its
implementation in \ba, is presented in Ref.~\cite{inprep}.
\section{Numerical results}
In order to check the technical implementation and the physical
accuracy of Eq.~(\ref{sigma}) in the new \ba~\cite{web},
the generator has been compared with independent
calculations, namely \labs~\cite{labspv},
\bh\ and the old $\tt 3.5$ release, tuning the relevant input parameters.
In \labs, which is not a true event generator, the corrected cross
section is calculated as $\sigma=(1+C_{NL})\sigma^\infty_{SF}$, where
$C_{NL}=(\sigma^{\alpha,ex}-\sigma^\alpha_{SF})/\sigma_0$ and the
Structure Functions in the strictly collinear limit are used: thus
\oal\ corrections are included exactly and h.o. are included in the
collinear limit with LL accuracy. \bh\ is an event generator based on the YFS
formalism to exponentiate exact \oal\ corrections. Here, \oal\ is
included exactly and h.o. are included in the YFS approach.
\begin{table*}[htb]
\caption{Comparison of \ba\ with \bh, \labs\ and the old version $\tt
 3.5$. Cross sections are in nb.}\label{tab1}
\begin{tabular}{ccccccc}
\hline
 setup& Born & \oal & \ba & \bh & \labs&\ba\ $\tt 3.5$\\
\hline
 a)& 6855.74(1)& 6059.9(1)& 6086.61(2)& 6086.3(2)& 6088.5(3)& 6107.7(2)\\
 b)& 529.463(1)& 451.53(1)& 455.853(4)& 455.73(1)& 456.20(1)& 458.44(1)\\
\hline
\end{tabular}
\end{table*}

The comparisons have been performed neglecting vacuum polarization
effects, at 1.02 GeV of center of mass energy, requiring for the final
state $e^+$ and $e^-$ $E_\pm>0.4\times E_{CoM}$, the $e^+e^-$
acollinearity $\xi$ lower than $10^{o}$ and choosing two angular
acceptances: a) $20^{o} < \vartheta_\pm < 160^{o}$ and b)~$55^{o} < \vartheta_\pm < 125^{o}$.

The results of the comparisons on the integrated cross section are
shown in Table~\ref{tab1}, where also the Born and \oal\ predictions are added.
In setup a) (b)), the \oal\ corrections reduce the Born cross section
by 13\% (17\%) and the corrections beyond \oal\ raise the cross
section by 0.4\% (0.9\%), showing that h.o. contributions are important at
the level of the needed accuracy. The agreement between \ba\ and \bh\
is very good, being the discrepancies well below the 0.1\%. The
tiny differences with respect to \labs\ (0.08\% in setup b)), where the
impact of
h.o. is larger) can be ascribed to the strictly collinear
approximation of h.o. corrections in \labs. Finally, the differences
between the new and old version of \ba\ are within the 0.5\%, in agreement
with the theoretical accuracy estimated for the old release
of the event generator.

\begin{figure}[htb]
\includegraphics[width=7.2cm]{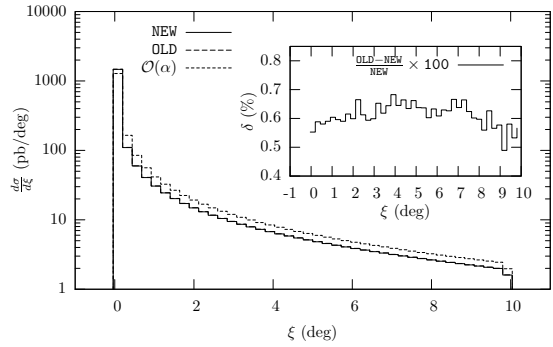}
\caption{Acollinearity distribution, the new \ba\ compared with its old version.}
\label{figacoll}
\end{figure}

Besides the integrated cross section, also differential cross sections
have been compared. In Fig.~\ref{figacoll}, the acollinearity
distribution is plotted for setup b), as obtained with the old and the
new \ba\ and at \oal. The effect of h.o. corrections is clearly
visible. The relative difference between the old and the new \ba\ is
better pointed out in the smaller panel, being almost constant in the
whole range at 0.5\% level.

\begin{figure}[htb]
\includegraphics[width=7.2cm]{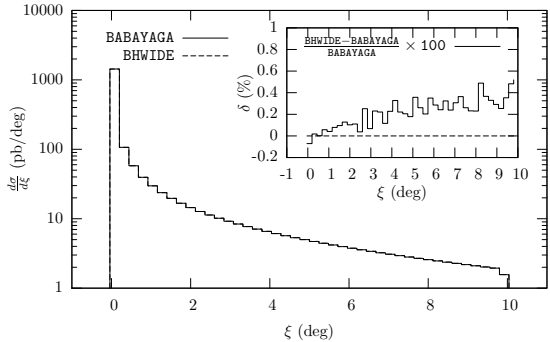}
\caption{Acollinearity distribution, \ba\ compared with \bh.}
\label{figacollbb}
\end{figure}

In Fig.~\ref{figacollbb}, the comparison of the acollinearity distribution
obtained with \ba\ and \bh\ is shown. The differences are very small
and can be only appreciated in the panel, showing they reach
the 0.4\% level only in the distribution tail. This region is
populated by events where hard photons are emitted and where the
different treatment of the multiple-photon emission and
h.o. corrections shows up, as can be expected.
\section{Estimate of the theoretical accuracy}
The results presented the previous section clearly demonstrate that
the matching
procedure of Eq.~(\ref{sigma}) has the desired behaviour of including
the missing \oal\ contributions in the PS algorithm while preserving
the resummation of all h.o. corrections. They also demonstrate that
the not trivial Monte Carlo implementation of Eq.~(\ref{sigma}) is
correctly realized and robust.

The theoretical error of \ba\ is now shifted at the two loop level, to
terms of the order of $\frac{1}{2}\alpha^2L$ ($\simeq 3\cdot10^{-4}$ at
1 GeV), without any infrared enhancement.

In~\cite{inprep}, a critical and systematic discussion of the remaining
theoretical error will be presented. Once the vacuum polarization
is (easily) included in the \ba\ formulation, the formulae can be expanded at
order $\alpha^2$ and consistently compared with the recent calculation
of Refs.~\cite{2loop}, where relevant contributions to the complete two loop
corrections to Bhabha scattering are computed. It is worth noticing
that in Refs.~\cite{2loop} the real photon radiation is
accounted only in the soft limit and that at two loop level also the
emission of real $e^+e^-$ pairs has to be carefully
considered. However, preliminary studies and
tests show that the missing ${\cal O}(\alpha^2)$ contributions have an
impact not larger than the 0.1\% on the integrated cross section if
typical event selection criteria for luminometry at flavour factories
are applied.

These considerations suggest that a conservative estimate of the
theoretical error of the improved \ba\ can be fixed at 0.2\%.
\section{Conclusions}
An improved version of the event generator \ba\ is now available~\cite{web},
where an original matching of the QED PS and the exact \oal\ matrix element is
implemented in order to go beyond the LL approximation intrinsic to the
PS approach. The \ba\ theoretical error for the Bhabha cross section
calculation at flavour factories is reduced from the 0.5\% down to
0.2\% (conservatively estimated).
A more detailed description of the
implementation and a more robust estimate of the theoretical accuracy
are discussed in Ref.~\cite{inprep}.
\section*{Acknowledgments}
\noindent
C.M.C.C. 
thanks the organizers of the 
workshop for the warm hospitality and the stimulating atmosphere
during the conference. The authors are grateful to
R.~Bonciani, A.~Denig, F.~Nguyen and G.~Venanzoni for useful discussions.

\end{document}